\documentclass[aip,cha,reprint,amsmath,amsfonts]{revtex4-1}
\usepackage[final]{graphicx}
\usepackage{dcolumn}
\usepackage{bm}

\usepackage{hyperref}
\hypersetup{
    colorlinks,
    citecolor=blue,
    filecolor=blue,
    linkcolor=blue,
    urlcolor=blue
}

\newtheorem{definition}{Definition}
\newtheorem{proposition}{Proposition}

\usepackage[inline,nomargin,author=]{fixme}

\makeatletter
\def\@lox@prtc{\section*{\@fxlistfixmename}\begingroup\def\@dotsep{4.5}}
\def\@lox@psttc{\endgroup}
\makeatother

\fxsetup{theme=color,mode=multiuser}
\fxsetface{inline}{\bfseries}

\definecolor{rorycol}{RGB}{200, 50, 0}
\definecolor{johncol}{RGB}{0, 200, 50}
\definecolor{kevincol}{RGB}{50, 0, 200}

\FXRegisterAuthor{ral}{aral}{\color{rorycol}RAL}
\FXRegisterAuthor{jrm}{ajrm}{\color{johncol}JRM}
\FXRegisterAuthor{kam}{akam}{\color{kevincol}KAM}
\providecommand{\e}[1]{\ensuremath{\times 10^{#1}}}

\begin{document}

\preprint{AIP/123-QED}

\title{Mode-locking in advection-reaction-diffusion systems: an invariant manifold perspective}

\author{Rory A. Locke}
\email{rlocke@ucmerced.edu}
\affiliation{School of Natural Sciences, University of California, Merced, California 95344, USA}
 
\author{John R. Mahoney}%
 \email{jrmahoney@ucdavis.edu.}
\affiliation{Department of Physics, University of California, Davis, California 95616, USA}%

\author{Kevin A. Mitchell}
  \email{kmitchell@ucmerced.edu.}
  \affiliation{School of Natural Sciences, University of California, Merced, California 95344, USA}%

\date{\today}

\begin{abstract}
Fronts propagating in two-dimensional \emph{advection-reaction-diffusion} (ARD) systems exhibit rich topological structure. When the underlying fluid flow is periodic in space and time, the reaction front can lock to the driving frequency. We explain this mode-locking phenomenon using so-called \emph{burning invariant manifolds} (BIMs).  In fact, the mode-locked profile is delineated by a BIM attached to a relative periodic orbit (RPO) of the front element dynamics. Changes in the type (and loss) of mode-locking can be understood in terms of local and global bifurcations of the RPOs and their BIMs. We illustrate these concepts numerically using a chain of alternating vortices in a channel geometry.
%
\end{abstract}

\maketitle

\listoffixmes

\begin{quotation}
For hundreds of years, Western travelers returning home from Southeast Asia told tales of fireflies perched in the branches and foliage along river banks flashing on and off in perfect unison. \cite{Strogatz:2003} Attempts to explain this synchrony included humidity, darkness, and observer error. How could these unintelligent insects achieve the same steady tempo we see from an orchestra of highly talented musicians? Perhaps there is one great maestro firefly leading the flashing ensemble? For years this seemed as plausible an explanation as any. It wasn't until the mid-1960s that experiments were performed to shed new light on the matter. 
Small groups of segregated fireflies, initially flashing randomly, gradually locked phases by speeding up or slowing down in response to their neighbors. This demonstrated their ability to adjust their flashing tempo to their external environment. Each firefly posses an internal metronome that is capable of changing tempo, thus giving them the ability to synchronize and bring order out of random flashing. Another interesting case of synchronization is called mode-locking. Instances of mode-locking have been observed in many dynamical systems including Josephson-Junctions \cite{V.-N.-Belykh:1977aa, Kawaguchi:2007aa}, lasers \cite{Haus:2000aa}, and musical instruments \cite{Mandal:2017aa,Fletcher:1999aa}. 
We are interested in mode-locking of chemical reactions in fluid flows, which occur when the natural (observed) frequency of the system becomes a rational multiple of the external oscillating frequency. Here we show how this mode-locking can be understood in terms of geometric structures called \emph{burning invariant manifolds}.

\end{quotation}

\section{\label{sec:level1}Introduction}

Many physical systems can be characterized by the propagation of a front. These systems range from plasmas\cite{D.-Beule:1998aa}, plankton blooms\cite{Scotti:2007aa}, the spread of disease, and chemical reactions\cite{Tel-T.:2005aa}. 
The case in which no flow exists, the reaction-diffusion limit, is well characterized by the theory of Fisher and Kolmogorov-Petrovskii-Piskunov (FKPP) \cite{A.-N.-Kolmogorov:1937aa}. The front velocity $v_{0}$, predicted by FKPP is given by:
 \begin{equation}
	v_{0}=2\sqrt{ \frac{D_{0}}{\tau}},
	\label{r9}	
\end{equation}
where $D_{0}$ is the molecular diffusivity and $\tau$ the reaction time-scale. 

However, analysis of fronts in flowing media have proved more challenging. In order to predict the front speed in a reacting fluid with flow, an attempt was made to modify FKPP theory by introducing an enhanced diffusivity, $D^{*}$ \cite{OMalley:2011}. This proved to be accurate only in the limit of a very slow reaction. Specifically, the enhanced diffusivity theory does not predict mode-locking in the periodically driven alternating vortex flow.
A specific type of mode-locking is specified by two integers $N$ and $M$.
\begin{definition}{$(N,M)$ \emph{mode-locking}}
is the recurrence of a pattern which has shifted by $N$ vortex pairs after having evolved under $M$ forcing periods.
\label{def1}
\end{definition}
It is straightforward to see that the average front velocity in the lab frame $v_f$ is related to the type of mode-locking through the following relation \cite{Cencini-M:2003aa}:
\begin{equation}
v_{f}= \frac{N\lambda}{MT},
\label{r10}
\end{equation}
where $\lambda$ is the width of a pair of vortices and $T$ the period of oscillation.

Cencini et.~al model ARD systems numerically using the so-called \emph{G}-equation in a grid-based approach\cite{Peters:2000aa,Cencini-M:2003aa,Abel:2002}. These methods do predict mode-locking, but are computationally costly\cite{Cencini-M:2003aa}. 
\begin{figure}
\includegraphics[width=\linewidth]{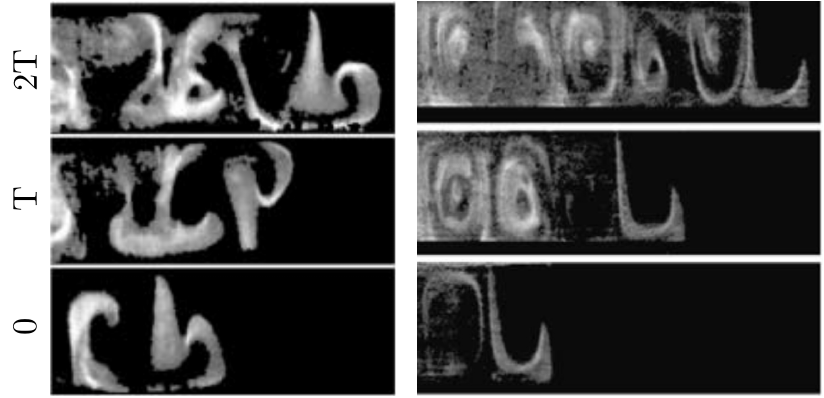}
\caption{\label{fig:ML} Experimental images of the Belousov-Zhabotinsky reaction in a quasi-two-dimensional alternating vortex channel flow. 
The flow is driven by an electric current passing through the fluid over a row of magnets.
Left: Experimental image of mode-locking type (1,2). 
It takes two driving periods for the ``b''-shaped front profile to repeat itself shifted by one pair of vortices.
Right: Experimental image of mode-locking type (1,1) \cite{Paoletti:2005aa}.}
\end{figure}
Several experiments have also clearly demonstrated the existence (and robustness) of mode-locking in ARD systems \cite{Paoletti:2005aa,Paoletti:2005ab}. Figure~\ref{fig:ML} shows such an experimental realization using the excitable Belousov-Zhabotinsky reaction in a chain of alternating vortices. 

Recently, it has been demonstrated that two-dimensional ARD flows with sharp reaction fronts can be reduced to a three-dimensional ODE for front-element dynamics \cite{John-Mahoney1a:2012aa, Mitchell:2012aa}. This approach reveals that reaction-front propagation is dominated by the presence of \emph{burning invariant manifolds} (BIMs), invariant manifolds of the front-element dynamics.
(We use the term ``burning'' as shorthand for any reaction front propagation differentiating them from their advective counterparts.)
Unlike traditional invariant manifolds, these BIMs act as \emph{one-way} barriers to reaction front propagation. In this article, we show how BIMs underly the phenomenon of mode locking. We explain how a BIM attached to a relative periodic orbit (RPO) of the front element dynamics can result in mode-locking and show numerical simulations. Finally we discuss how changes in the type, as well as the loss, of mode locking can be understood in terms of local and global bifurcations of the RPOs and their BIMs.

This paper is organized as follows.
 Section~\ref{sec:level2a} introduces the three-dimensional dynamical system for a point along the front. 
Section~\ref{sec:level2b} reviews the theory of burning invariant manifolds.  
Section~\ref{sec:level3a} justifies the abstract theory with concrete numerical realizations of this connection in a model flow.
Section~\ref{sec:level3b} presents the central result of this paper, establishing the fundamental connection between mode-locking and BIMs.
Finally, Section~\ref{sec:level4}discusses bifurcations in mode-locking type as a function of the front propagation speed and how global bifurcations can create or destroy mode locking of a given type .
Our numerical technique for locating RPOs is discussed in the appendix.

\section{Preliminaries}
	
\subsection{Front-element dynamics}

\label{sec:level2a} 

We model ARD systems by considering only the reaction front. This is more computationally efficient than explicitly modeling the entire fluid state and, we believe, more theoretically insightful. We make use of the following assumptions. First, the reaction time-scale is much smaller than the diffusion time-scale---this is known as the ``sharp front'' or geometric-optics limit \cite{Abel:2001,Abel:2002}. 
In this limit, the fluid is divided into reacted and unreacted regions that are separated by a well-defined boundary---the reaction front. Second, we assume that each front element progresses in a manner independent of the local curvature of the front. It is known that this curvature plays an important role in certain systems \cite{Neufeld:2009aa}. Last, we assume that, in the comoving fluid frame, the front propagates with a fixed speed $v_{0}$ that is homogenous and isotropic. 

More technically, a front is defined as the oriented boundary between reacted and unreacted regions where the local orientation vector $\hat{\mathbf{n}}$ is normal to the reaction front and points \emph{away} from the burned region. The orientation might also be specified by the tangent vector $\mathbf{\hat{\mathbf{g}}}$, where $\mathbf{\hat{\mathbf{g}}}$ is orthogonal to $\mathbf{\hat{\mathbf{n}}}$ and $\mathbf{\hat{\mathbf{n}}} \times\mathbf{\hat{\mathbf{g}}} = +1$.  If $\mathbf{r}$ is used to denote the $xy$-position of a front element and $\theta$ as the angle between the positive $x$-axis and $\mathbf{\hat{\mathbf{g}}}$, a front in three-dimensional $xy\theta$-space is a curve $(\mathbf{r}(\lambda), \theta(\lambda))$ that satisfies the \emph{front-compatibility criterion}
	\begin{equation} \label{eq.1}
	 \mathbf{\frac{\mathrm{d}r}{\mathrm{d}\lambda } = \hat{g}(\theta)},
	\end{equation}	
where $\lambda$ is the Euclidean length parameter measured in $xy$-space that increases in the $+ \mathbf{\hat{\mathbf{g}}}$ direction. 

Each front element, specified by $(\mathbf{r}(t), \theta(t))$, evolves via
\begin{subequations}
\begin{align} 
	\dot{\mathbf{r}} &= \mathbf{u} + v_{0}\hat{\mathbf{n}}, \label{eq.4a} \\
	\dot{\mathbf{\theta}}& = -\sum_{i,j} \hat{n}_i u_{i,j}\hat{g}_j,
	\label{eq.4b}
\end{align}
\label{r1}
\end{subequations} 
where $\mathbf{u}$ is the incompressible fluid velocity and $u_{i,j} =  \partial u_i/ \partial r_j $.
 Each fluid element is translated by the vector sum of the fluid velocity and the burning velocity, Eq.~(\ref{eq.4a}). Furthermore, the change in orientation is determined solely by the local behavior of the fluid flow, Eq.~(\ref{eq.4b}). Numerically, a front is composed of a line of discrete fluid elements that propagate independently under this three-dimensional ODE. To maintain a sufficiently smooth front in spite of the stretching of the fluid flow, we dynamically insert new points by interpolation.

\begin{figure}
\includegraphics[width=\linewidth]{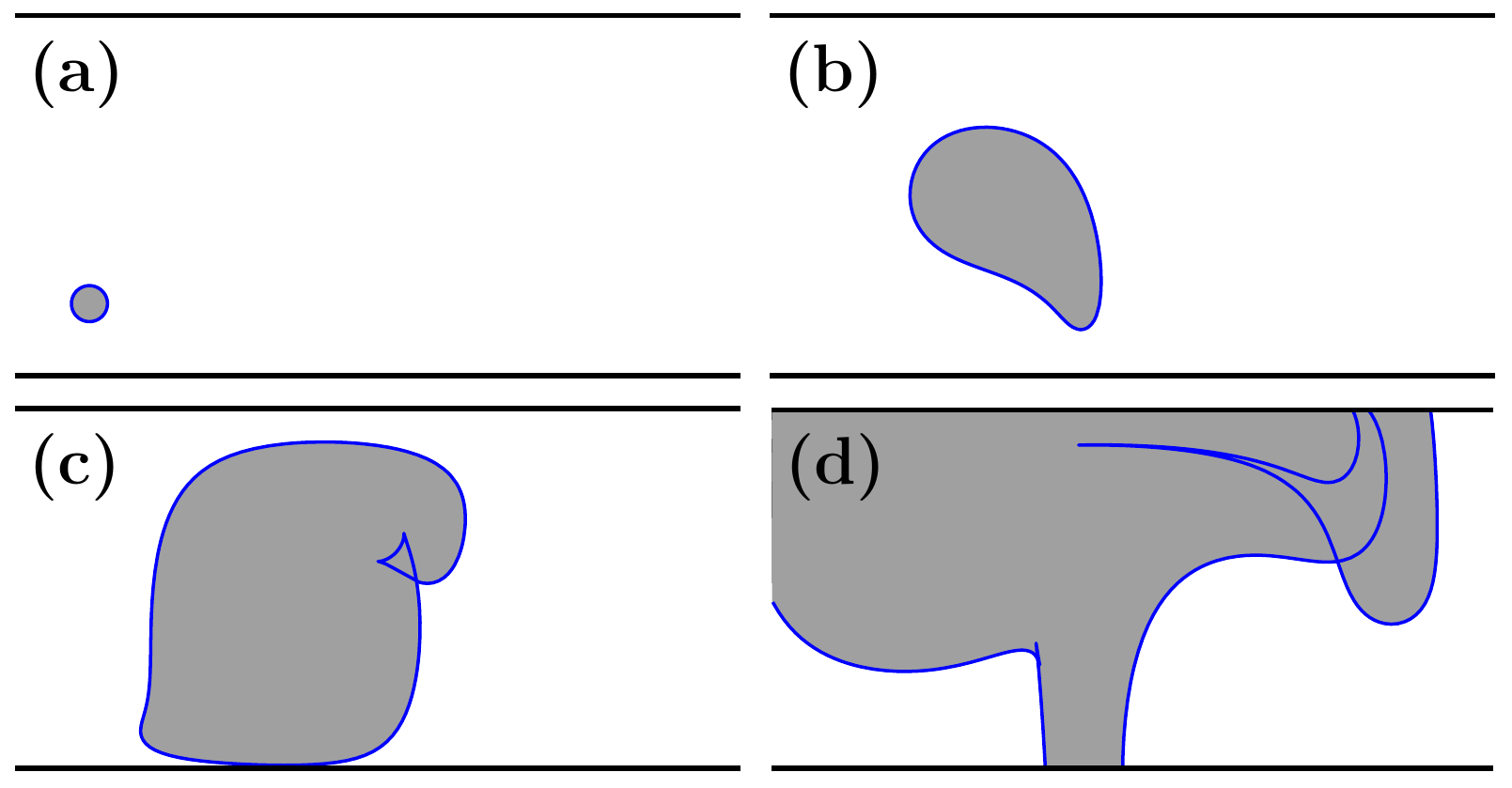}
\caption{\label{fig:bounding}
(a) An initial circular burned region at $t = 0T$. Panels (b)-(d) show the forward evolution at times $t = 0.3 T, 0.6 T,$ and $1.19 T$ respectively. The front is evolved using Eq.~(\ref{r1}) with $v_{0}$= 0.2 and $\mathbf{u}$ given by Eq.~(\ref{r2}), with $b = 0.25$, and $\omega = 2$.}
\end{figure}

As it evolves, portions of the front may come to lie within the burned region. 
Figure~\ref{fig:bounding} shows how an initially circular front (a) grows into a non-circular front (b), develops a swallowtail (c), and encounters the channel boundary (d).
The swallowtail portion of the front lies within the burned region and therefore no longer represents a physical boundary. While we refer to the entire evolved curve (either in $xy$-space or $xy\theta$-space) as the front, we refer to those segments of the front separating the burned from unburned regions as the \emph{bounding front}. 
The bounding front can exhibit corners where the bounding front is not differentiable in $xy$-space (i.e., it is disconnected in $xy\theta$-space).

\subsection{Burning invariant manifolds (BIMs) and frozen fronts}
\label{sec:level2b} 

The theory of invariant manifolds in passive advection is well established.
In the two-dimensional setting, time-independent flows lead to separatrices---invariant manifolds that connect various fixed points and divide the fluid into ``cells''.
Under a time-periodic perturbation, these separatrices split giving rise to distinct stable and unstable manifolds.
These more complicated objects describe the mechanism of transport between cells.

The addition of propagation dynamics has several interesting consequences.
First, we find fixed points of Eq.~(\ref{r1}) where the front propagation exactly counters the fluid flow.
We refer to these as \emph{burning} fixed points.
Of primary interest are the stable-stable-unstable (SSU) burning fixed points.
Attached to these are the so-called \emph{burning} invariant manifolds (BIMs).
Here we focus on the 1D unstable BIMs attached to SSU burning fixed points. 
It has been shown that BIMs serve as \emph{one-way} barriers to front propagation\cite{John-Mahoney1a:2012aa, Mitchell:2012aa}. 
The bounding behavior of BIMs is due to the ``no-passing'' lemma: no front can overtake another front from behind. 
Thus, fronts oriented in the same direction as the BIM will be unable to directly pass through. 
Conversely, fronts oriented opposite the BIM will pass through unobstructed.
Also, arbitrary fronts (within some moderate basin of attraction) converge to the BIMs.
Because of these facts, SSU BIMs are important for understanding the evolution and long-time behavior of ARD systems.

Another new property of BIMs is a phenomenon more familiar in optics:
BIMs can form cusps and self-intersections when projected onto the \emph{xy}-plane as shown in Fig.~\ref{fig:bounding}(c).
(In the full \emph{xy$\theta$}-phase space, BIMs do not self-intersect due to the uniqueness of solutions to ODEs.) 
In time-independent flows, a cusp marks the end of the physically relevant bounding behavior of the BIM. 

Finally, as first observed experimentally just over a decade ago, reaction fronts in steady ARD flows exhibit the tendency to pin to vortex structures in the presence of an imposed ``wind'' \cite{Megson:2015aa, John-R.-Mahoney:2015aa}. 
A front initiated upwind of a pinning site can travel downstream, catch the site, and eventually reach a steady state.
This ``frozen front'' phenomenon has been explained in terms of BIMs.
Qualitatively, a \emph{frozen front} occurs when a BIM spans the entire channel without a cusp, or when a set of BIMs do so collectively. These results can be made precise using the concept of a BIM core; the BIM core is defined as the BIM segment that includes the burning fixed point and extends in both directions until reaching either a cusp, a new burning fixed point, or infinity.

\begin{proposition}
\label{prop:frozen_front}
Frozen fronts in steady flows are built from BIM cores. More precisely, each frozen front is generated by a set of SSU burning fixed points. The frozen front is obtained by tracing the unstable manifold from each point in the set of fixed points until one of three things occurs: it intersects any other BIM core emanating from this set; it intersects any domain boundary; or it terminates at an SSS burning fixed point \cite{John-R.-Mahoney:2015aa}.\end{proposition}

For time-periodic flows, the bounding behavior of BIMs is more subtle.
These BIMs can stretch and fold in time (with plenty of cusps and self-intersections) allowing the reaction front to propagate down the channel via a turnstile-like mechanism \cite{Mahoney:2016aa,Mackay:1983} .
In spite of the apparent loss of bounding behavior (in the lab frame), time-periodic driving yields a new form of stable structure in a moving frame---mode-locked fronts.

\section{Mode-Locking} 

\subsection{Numerical Simulations}
\label{sec:level3a}

As a concrete example we choose the well-studied alternating vortex chain {\cite{Chandresekhar:1961aa,Solomon:1988aa,Cencini-M:2003aa}. This flow mimics a two-dimensional cross-section of Rayleigh-Benard convection in which roll patterns, i.e. vortices, appear due to an instability driven by heating of the lower boundary. The two-dimensional velocity field is given by
\begin{subequations}
  \begin{align} 
    u_{x}(x,y,t) = +\sin(\pi[x+b\sin(\omega t)])\cos(\pi y), \label{eq.1}	\\
    u_{y}(x,y,t) = -\cos(\pi[x+b\sin(\omega t)])\sin(\pi y), \label{eq.2}	
  \end{align}
  \label{r2}
  \end{subequations}
where $0\leq y \leq 1$ and time-dependence is produced by the lateral oscillation term $x+b\sin(\omega t)$. Note that this model assumes free-slip boundary conditions. The dimentionfull parameters of the system are $U$,$D$, $\Omega$, $B$ and $V_{0}$ which correspond to the maximum fluid speed, channel width, driving frequency, driving amplitude, and the front propagation speed in the absence of a flow, respectively. The dimensionless parameters of the system are $b = B/D$, $\omega = \Omega D/U$, and $v_{0} = V_{0}/U$. The position vector \textbf{r} is scaled such that the maximum fluid vortex speed, \emph{U}, is one and the width of each vortex is one. For these simulations we fix $b = 0.3$, $\omega = 4.08$ and vary the front velocity $v_{0}$. Numerical simulations were performed by propagating individual front elements under Eq.~(\ref{r1}). 
By varying parameters, we identified mode-locking of types: (1,1), (1,2), (2,3) and (3,5).

\begin{figure}
\includegraphics[width=\linewidth]{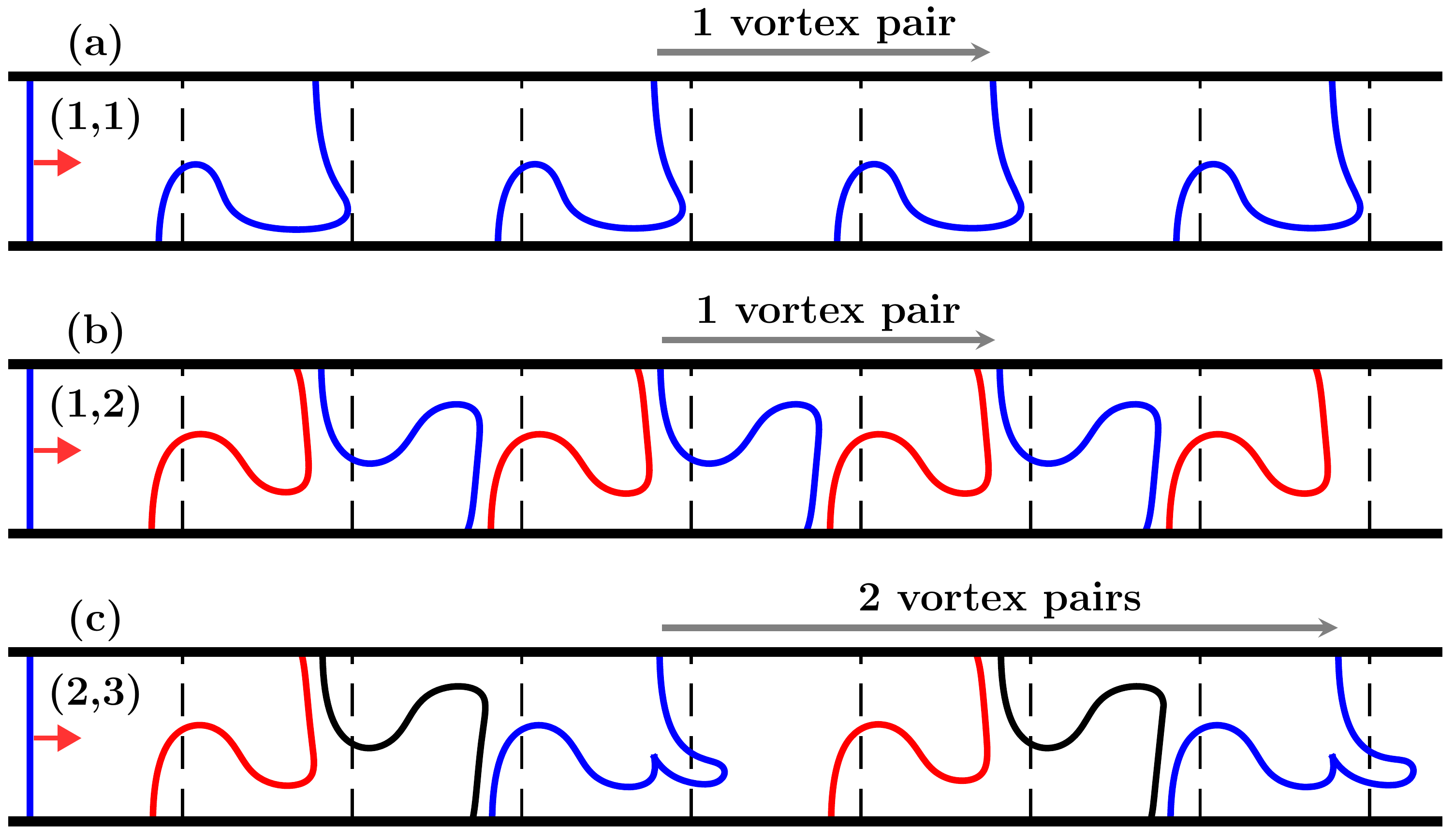}
\caption{
Numerically generated reaction front showing evidence of mode-locking. The front propagates down the channel from left to right starting with a vertical front spanning the channel. After one period the front has fully converged. Fig.~\ref{fig:ML_all} (a)-(c) shows mode-locking of types (1,1), (1,2), and (2,3)  with $v_{0}$ = 0.18, 0.068, and 0.089 respectively. Dashed vertical lines indicate the vortex cells.}
\label{fig:ML_all}
\end{figure}

Figure~\ref{fig:ML_all}a shows snapshots of a front propagating down the channel from left to right, beginning with a vertical front that spans the channel. Each snapshot is taken after one driving period.  We see that the front converges almost instantaneously (essentially after the first iterate) to a mode-locked front: the front profile repeats after $M = 1$ driving period and is translated to the right by $N = 1$ vortex pair, demonstrating type (1,1) mode-locking.  Similarly, in Fig.~\ref{fig:ML_all}b the front profile repeats after $M = 2$ driving periods, translated by $N = 1$ vortex pair for type (1,2) mode-locking.  Finally, Fig.~\ref{fig:ML_all}c shows type (2,3) mode-locking.

\begin{figure}
\includegraphics[width=\linewidth]{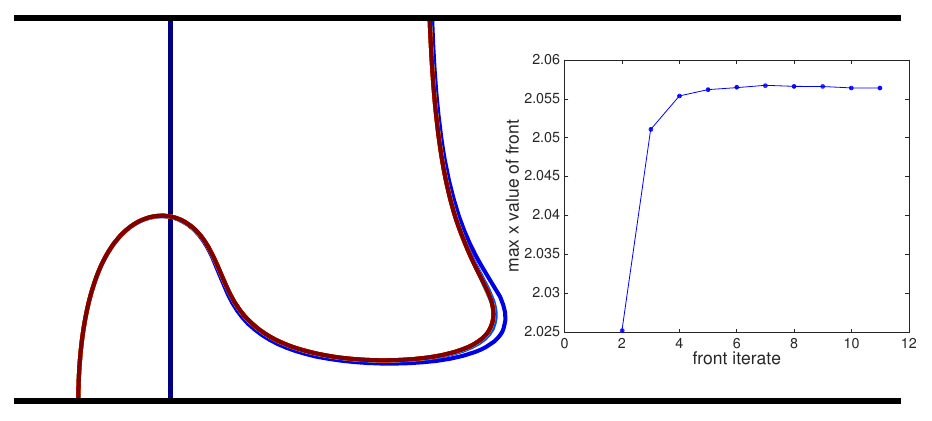}
\caption{Simulation demonstrating convergence of fronts to mode locking. (1,1) mode-locking with fronts shifted backwards showing convergence to be approximately one period. The initial condition is shown as the straight vertical line. The inset shows the rightmost point of the front as a function of the number of iterates. Beginning with the initial front we see it takes no more than two iterates to reach the mode-locked state.} 
\label{fig:Front_Convergence}
\end{figure}

The convergence of the initial vertical front to the mode-locked profile is very rapid.  Figure ~\ref{fig:Front_Convergence} shows the initial condition as the vertical line.  Each of the subsequent fronts in Fig.~\ref{fig:Front_Convergence} is shifted backward by an integer number of vortex pairs, so that the fronts all lie within the original cell.  Except for the initial vertical line, these shifted fronts are visually nearly identical, demonstrating that a single iterate is essentially all that is needed to reach the mode-locked state.  

\begin{figure}
\includegraphics[width=\linewidth]{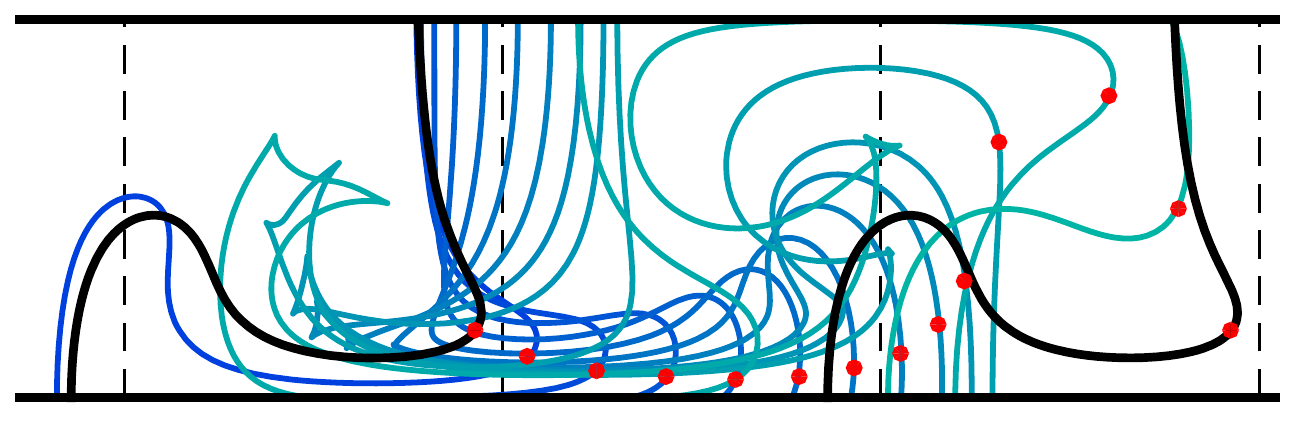}
\caption{Reaction front shows evidence of (1,1) mode locking with $v_{0} = 0.18$. The front evolves from left to right down the channel beginning with a fully converged front, black curve on the left, and then evolving forward over one period where the shape of the front repeats itself one period later, black curve on the right.  
}
\label{fig:continuous}
\end{figure}
After a front has converged to the mode-locked profile, Fig.~\ref{fig:continuous} shows several equally spaced intermediate time steps of its evolution over one mode-locked period for the (1,1) mode-locking case. The black curve on the left maps to the black curve on the right one mode locked period later. The intermediate curves are snapshots of how a front evolves over one period. Note that the propagation of the front down the channel is not strictly monotonic.  As Fig.~\ref{fig:continuous} shows some intermediate steps are closer together and some farther apart. Also it is interesting to see the formation and loss of swallowtails in between periods as the front evolves. Lastly we could chose any intermediate front and map it forward one period and that shape would repeat, there is nothing special about the shape of the front in respect to mode-locking.
\begin{figure}
\includegraphics[width=\linewidth]{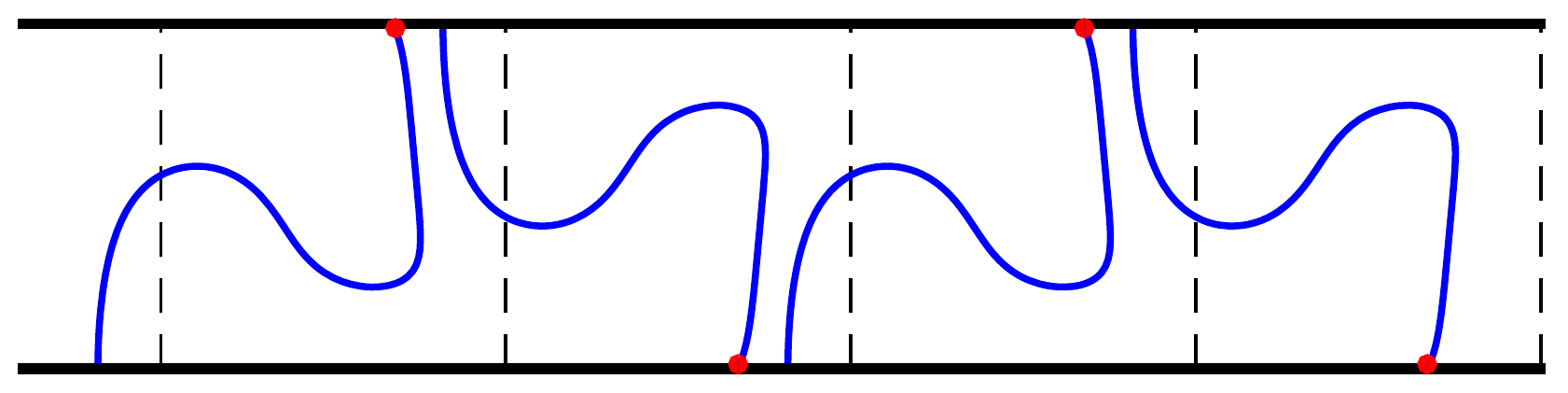}
\caption{ Simulation of (1,2) mode-locked front propagating down the channel for two periods. 
}
\label{fig:12ML}
\end{figure}

\begin{figure*}
\includegraphics[width=\linewidth]{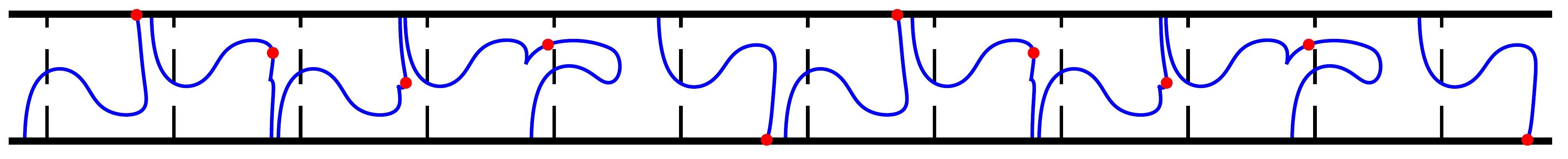}
\caption{Simulation of (3,5) mode-locked front propagating down the channel for nine periods. 
}
\label{fig:35ML}
\end{figure*}

Figure ~\ref{fig:12ML} and Figure ~\ref{fig:35ML} show mode-locked fronts  of type (1,2) and (3,5) respectively. These images show that the BIM model can also accurately capture higher order, i.e. period, mode-locking similar to the grid based approach used by Cencini et. al. "Finding" higher order mode-locking presents some interesting numerical challenges however, we developed a robust approach that will be covered in detail in the appendix. 

\subsection{Mode-locked fronts are composed of BIMs} 

\label{sec:level3b}

In this section, we derive our main mathematical result, Proposition~\ref{prop:MLBIMs}, a structural characterization of mode-locked fronts. We first introduce some notation. The stroboscopic map $F$ evolves an initial point $(x,y,\theta)$ forward via Eq.~(\ref{r1}) for one forcing period $T$.
The shift map $S$ translates a point forward by a single vortex pair, i.e.
\begin{equation}
S(x,y,\theta) \equiv (x + 2, y, \theta).
\label{r3}
\end{equation}
Finally, we define the composite map $F^{(N,M)}$,
\begin{equation}
F^{(N,M)} \equiv S^{-N} \circ F^{M},
\label{r4}
\end{equation}
which simply evolves an initial point forward $M$ forcing cycles and then shifts it backward by $N$ vortex pairs.

\begin{figure}
\includegraphics[width = \linewidth]{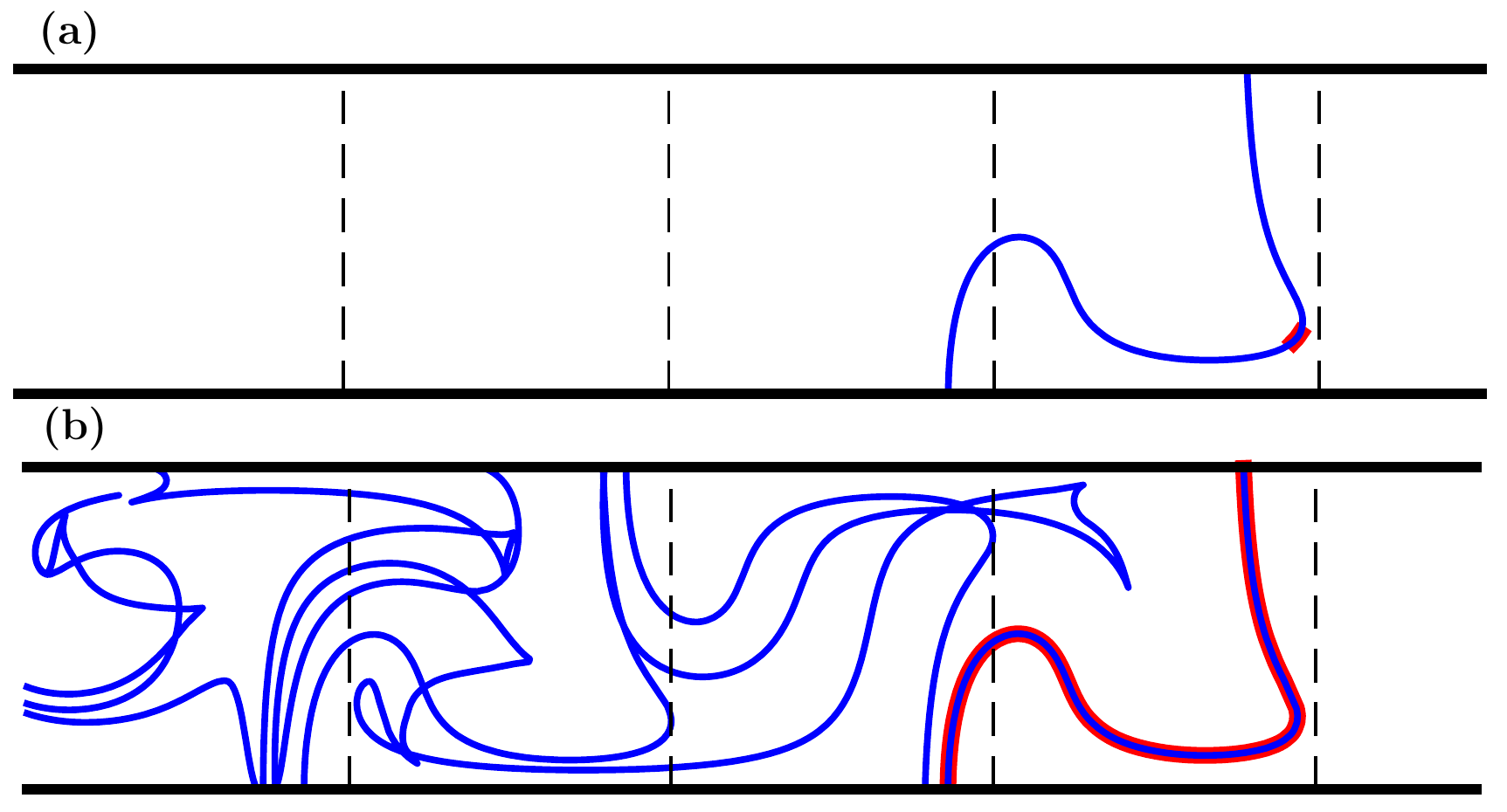} 
\caption{
\label{fig:stretching_BF} Simulation of (1,1) mode-locking demonstrates the extreme stretching experienced by the bounding front under the map $F^{(1,1)}$.
Blue bounding front (a) maps to blue front (b).
Small red segment (a) maps to red bounding front (b).
}
\end{figure}
By definition, an $(N,M)$ mode-locked front must be invariant under $F^{(N,M)}$ in order to satisfy the original characterization of mode-locking, Definition~\ref{def1}.  
Importantly, the mode-locked front is on the boundary between the burned and unburned regions, i.e. it is the bounding front $B$ of the potentially longer curve that may extend into the burned region.
As with frozen fronts in stationary flows, the projection of $B$ into $xy$-space need not be a single smooth segment, but may have corners where two segments join.  
Unlike frozen fronts, however, the projection of $B$ may have multiple disconnected pieces.  
One of these pieces constitutes the \emph{leading front} of the mode-locked front, i.e., the part of the curve that directly contacts the infinite unburned domain.
This leading front consists of a finite union of smooth segments connecting the top of the channel to the bottom. 
The other connected components of the mode-locked front all bound voids that trail the leading front.  

Note that in the full phase space, $B$ is a finite union of closed disjoint segments.  As $B$ evolves forward in time, these segments will be stretched and folded so that they are ultimately longer than the original $B$.  Thus, $B$ is only invariant in the sense that $F^{(N,M)}(B)$ includes the original $B$, i.e. $B \subset F^{(N,M)}(B)$.  Figure~\ref{fig:stretching_BF} shows an example of this behavior, where an initial bounding front $B$ (blue in Fig.~\ref{fig:stretching_BF}a) evolves forward to the curve $F^{(N,M)}(B)$ (blue in Fig.~\ref{fig:stretching_BF}b).
Because of the extreme stretching, the tiny red segment in Fig.~\ref{fig:stretching_BF}a grows to cover the entire leading front (red) in Fig.~\ref{fig:stretching_BF}b.

We next show that for the system to exhibit mode-locking of type $(N,M)$, there must exist a \emph{relative periodic orbit} (RPO) of Eq.~(\ref{r1}) on the mode-locked front $B$.  A relative periodic orbit is an orbit that is periodic when viewed relative to the comoving reference frame, i.e. the reference frame moving with a uniform velocity equal to the average front velocity $v_f$ [Eq.~(\ref{r10})].  More precisely, we define a relative periodic orbit of type $(N,M)$ to be an orbit that is shifted forward $N$ vortex pairs after $M$ driving periods, which can be equivalently stated as follows.
\begin{definition}
An $(N,M)$ relative periodic orbit (RPO) is a fixed point under $F^{(N,M)}$. 
\end{definition}
Much of the structure of the mode-locked front $B$ is revealed by analyzing the inverse map $(F^{(N,M)})^{-1}$, which shrinks $B$, i.e. $(F^{(N,M)})^{-1}(B) \subset B$.  Since $B$ is a collection of segments, $(F^{(N,M)})^{-1}$ restricted to $B$ is topologically equivalent to a differentiable map $f:\cup_i I_i \rightarrow \cup_i I_i$ defined over a finite collection of disjoint closed intervals $\{I_i\}$ on the real line.  For any such $f$, there must exist some interval $I'$ that maps into itself after some number of iterates $k \ge 1$, i.e. $f^k(I') \subset I'$.  By the Brouwer Fixed Point Theorem~\cite{Munkres:1993aa} (equivalently, the Intermediate Value Theorem in 1D) $f^k$ has a fixed point $x\in I'$.  More generally, it is easy to see that $f^k$ will generically have a finite odd number of fixed points in $I'$ that alternate between stable and unstable, i.e. SUSU...US, with at least one stable fixed point.  Thus, all points in $I'$ (except the unstable fixed points) will converge to a stable fixed point under $f^k$.  These facts are true for any interval $I'$ that eventually maps into itself under $f$.  For any interval $\tilde{I}$ that doesn't eventually map into itself, $\tilde{I}$ will eventually map into an interval that does, i.e. $f^\ell(\tilde{I}) \subset I'$, and hence all points in $\tilde{I}$ converge to a stable fixed point of $f^k$.  These results for $f$ readily imply the following facts for the mode-locked front $B$.

\begin{figure}
\includegraphics[width = \linewidth]{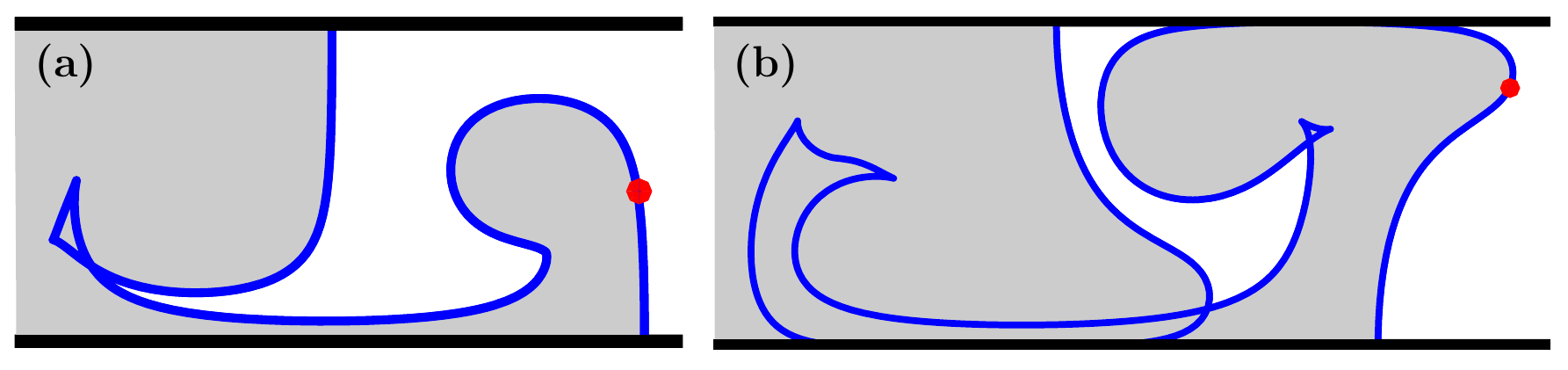} 
\caption{(a) Mode-locked front with RPO in red. The front has a swallowtail that extends into the burned region. Here the bounding front is composed of two segments only one of which has the RPO. (b) Mode-locked front with RPO in red, and a void (unburned region) trapped behind the bounding front. If an RPO of another type were present on any segment of the front that encompasses the void it would not persist indefinitely. All the unburned material behind the bounding front will become burned thus it is not possible to have two simultaneous mode-locking types.}
\label{fig:Swallowtail_RPO}
\end{figure}

(i) A mode-locked front of type $(N,M)$ contains an RPO of type $(k N,k M)$ for some integer $k \ge 1$.  This RPO is unstable in the direction tangent to the front.  Furthermore, each smooth segment of the mode-locked front either has no RPOs or a finite odd number of RPOs that alternate between unstable and stable in the direction tangent to the front, i.e. USU...SU.  Thus, while each smooth segment of a frozen front must contain a BFP (Prop.~\ref{prop:frozen_front}), it is not necessary for each smooth segment of a mode-locked front to contain an RPO.  For example, the mode-locked front in Figure~\ref{fig:Swallowtail_RPO}a has two smooth segments joined at a swallowtail.  Only one of these segments contains an RPO.  

(ii) Every point on the mode-locked front (except the RPOs that are stable in the tangent direction) is in the unstable manifold of an RPO that is unstable in the tangent direction.  Note that the smooth segment in Fig.~\ref{fig:Swallowtail_RPO}a that does not contain an RPO nevertheless lies within the unstable manifold of the RPO.

Result (i) above can be further refined: only the leading front can contain RPOs.  This is because each void trailing the leading front can only persist for a finite amount of time, as the void is of finite area and the front consumes unburned fluid at a constant rate per unit length.  The unburned fluid inside a void must be repeatedly refreshed by the voids being ``pinched off'' from the leading front, with the boundaries of the voids thus being on the unstable manifolds of RPOs on the leading front.  This is exactly what is seen in Fig.~\ref{fig:Swallowtail_RPO}b.  Furthermore, if there are multiple RPOs (of type $(kN, kM)$) on the leading front, their ordering along the front cannot change under $F^{(N,M)}$; if this were to happen then one RPO would have to pass in front of the other and the trailing RPO would end up in a void or within the burned region. Neither of these is possible.  Thus, no RPO can trade places with any other RPO under the map $F^{(N,M)}$, which means that each RPO is a fixed point of $F^{(N,M)}$ so $k = 1$.

We now consider only those mode-locked fronts that are stable to small perturbations, making them physically relevant to the experiments.  This stability of the front implies that all RPOs must be stable in the remaining two directions, i.e. the RPOs have stability SSU or SSS in the full phase space.

Thus, the mode-locked front is generated by a finite number of SSU RPOs, where the front continuously grows out of the RPOs as time evolves.  Tracking a point on one of these BIMs as a function of time, we find that it ceases to be on the mode-locked front only when it strikes the channel wall or hits another point along the BIM, where it then penetrates the burned region.  A point on the mode-locked front can also converge onto an SSS RPO on the front.  We summarize with the following proposition analogous to Proposition~\ref{prop:frozen_front}.

\begin{proposition}
\label{prop:MLBIMs}
Mode-locked fronts are built from BIMs.  More precisely, a physically stable mode-locked front of type $(N,M)$ is generated by a set of SSU RPOs of type $(N,M)$.  The mode-locked front is obtained from those points on the BIMs of the RPOs that have never previously struck some other point on the BIMs or struck the channel wall.  The BIM segments in the mode-locked front may have endpoints at the wall, at another BIM segment, or at an SSS RPO.
\end{proposition}

Finally, it is important to note that the RPOs that generate the front must in fact lie on the bounding part of their BIMs.  That is, as the BIMs grow out from the RPOs, they must not burn through an RPO.  Said another way, each RPO must never pass through its own BIM or that of another SSU RPO.  This possibility will be discussed in greater detail in Sect.~\ref{sec:level4}.

\section{\label{sec:level4}Changes in mode-locking as $v_{0}$ is increased.}

\begin{figure}
\includegraphics[width=\linewidth]{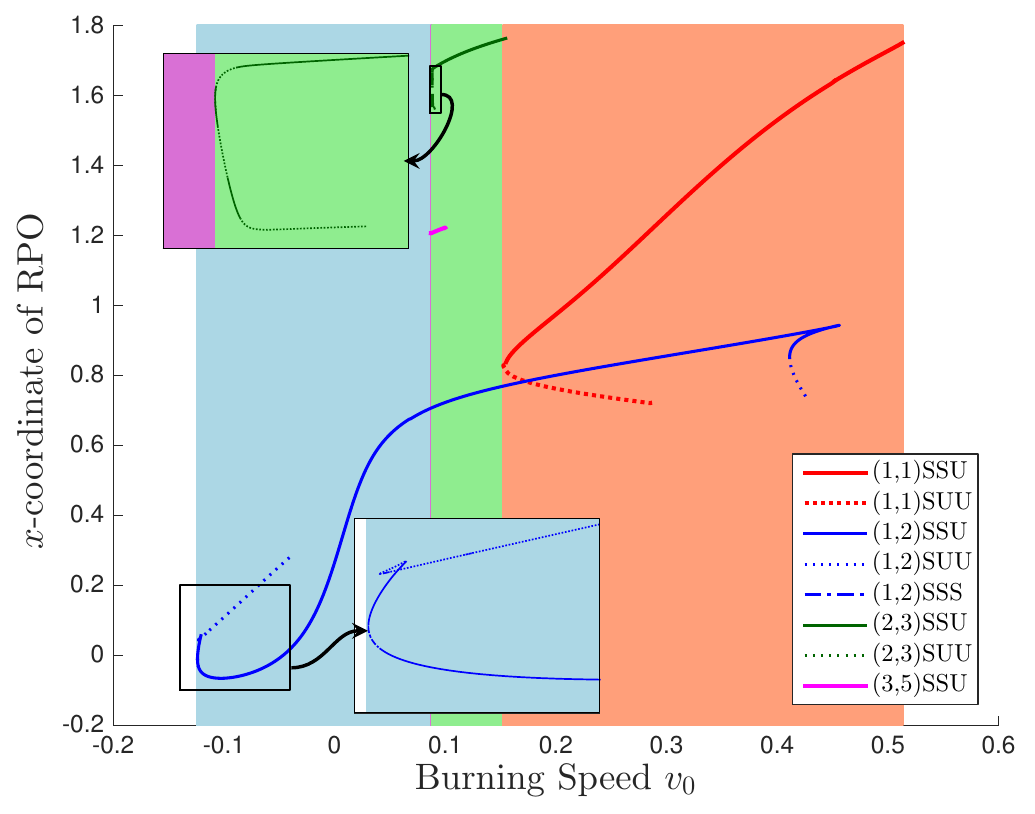}
\caption{
Bifurcation diagram of the RPOs as a function of $v_0$.
Multiple RPOs exist for the same parameters.
Average front speed of dominant ML front (Eq.~\ref{r10}) is monotonic in $v_0$.
Insets shows that each ML RPO is created in a saddle-node bifurcation.
}
\label{fig:Continuation}
\end{figure}

The dynamical systems approach outlined in Sect.~\ref{sec:level3b} leads to a greater understanding of the underlying mechanism for mode-locking. Figure~\ref{fig:Continuation} is a plot of the $x$-coordinate of one point of an RPO as a function of front speed $v_{0}$.  Here we show SSU and SUU RPOs for mode-locking types (1,1), (1,2), (2,3), and (3,5). Following the burning speed from left to right we see that mode-locking orbits are born in saddle-node bifurcations. 

Figure~\ref{fig:Continuation} also shows that for a given value of $v_{0}$, multiple RPOs can coexist. The mode-locking type that is physically realized will be the type with the fastest RPO. This must be true because the mode-locked front with the fastest burning speed will overtake any other RPOs present engulfing them in the burned region.
Note that while the overtaken RPO may temporarily exist within a void, this void will vanish in a finite time as argued previously.
Thus it is not possible to have two simultaneous mode-locking types. 
We refer to the fastest mode-locking RPO (or front) as being \emph{dominant}.

Observe that the mode-locking front speed increases monotonically with $v_{0}$ as can be seen from the progression of mode-locked types from (1,2) $\to$ (3,5) $\to$ (2,3) $\to$ (1,1), with speeds in proportion to $1/2 \to 3/5 \to 2/3 \to 1$. To explain this monotonicity consider the evolution of a mode-locked front with parameter $v_0$. Now consider the evolution of the same initial front with parameter $v_0 + \epsilon$. 
The region burned by the second front will be a superset of the region burned by the original mode-locked front.
Therefore the front with the larger $v_0$ cannot travel more slowly. 

Notice that the loss of a particular mode locking type $(N,M)$ is not due to the disappearance of the $(N,M)$ RPO.
For example, in Fig.~\ref{fig:Continuation}, the (1,2) SSU RPO disappears in a saddle-node bifurcation at $v_0 \approx 0.45$.
However, by this value of $v_0$, the (1,1) RPO has already taken over the dominant mode-locking role.
In fact, the mode-locking role is lost even earlier, by $v_0 \approx 0.09$ when the (3,5) RPO is created.
The question remains: what has changed about the structure of the (1,2) SSU RPO and its unstable manifold that causes it to lose its dominance?
Clearly, it will lose dominance as soon as a faster RPO is created.
But, can we determine when dominance is lost by looking at the RPO and its unstable manifold alone?
We shall see below that loss of dominance coincides with a global bifurcation in the structure of the unstable manifold.

Figure~\ref{fig:Global Bifurcation}a shows a $(1,2)$ RPO and a piece of its BIM at time $t = 0$, with burning speed $v_{0} = 0.0873$.  Using the data from Fig.~\ref{fig:Continuation}, it can be shown that there is also a mode-locked front of type $(3,5)$ at this value of $v_0$.  Because the $(3,5)$ RPO is faster, we can conclude that the $(1,2)$ RPO cannot be dominant. What about the structure of the $(1,2)$ front causes it to lose its dominance? To appreciate the mechanism at work, we evolve the $(1,2)$ front forward one period in Figure~\ref{fig:Global Bifurcation}.  Over that time, the RPO moves upward from the bottom boundary to the top boundary.  At the same time, the BIM develops a long ``finger" along the top boundary, Figs.~\ref{fig:Global Bifurcation}b and \ref{fig:Global Bifurcation}c.  As the RPO continues upward it collides with the finger; importantly, it may then pass into the finger because the colliding fronts have opposite burning directions, Fig.~\ref{fig:Global Bifurcation}d.  Finally, at the end of one period, the RPO is no longer on the bounding front.  As discussed in Sect.~\ref{sec:level3b}, a physical mode-locked front must be the boundary between the burned and unburned regions, and thus this RPO fails to generate a dominant mode-locked front.  In other words, an RPO fails to be dominant if it passes through its own BIM or the BIM of another mode-locked RPO.  This is already implied in Proposition~\ref{prop:MLBIMs} by the statement: ``The mode-locked front is obtained from those points on the BIMs of the RPOs that have never previously struck some other point on the BIMs.''  The reason Fig.~\ref{fig:Global Bifurcation}a is deceptive is because we have not plotted a sufficiently long piece of the BIM.  A necessary condition, and a nice check, for whether an RPO is dominant is that the evolution of the front over one period does not produce any new segments on the bounding front, which is clearly violated in Fig.~\ref{fig:Global Bifurcation}.  Finally, when varying $v_0$, a dominant mode-locked front loses its dominance at that value of $v_0$ where the RPO first lies on (the $xy$-projection of) the BIM itself.

\begin{figure}
\includegraphics[width=\linewidth]{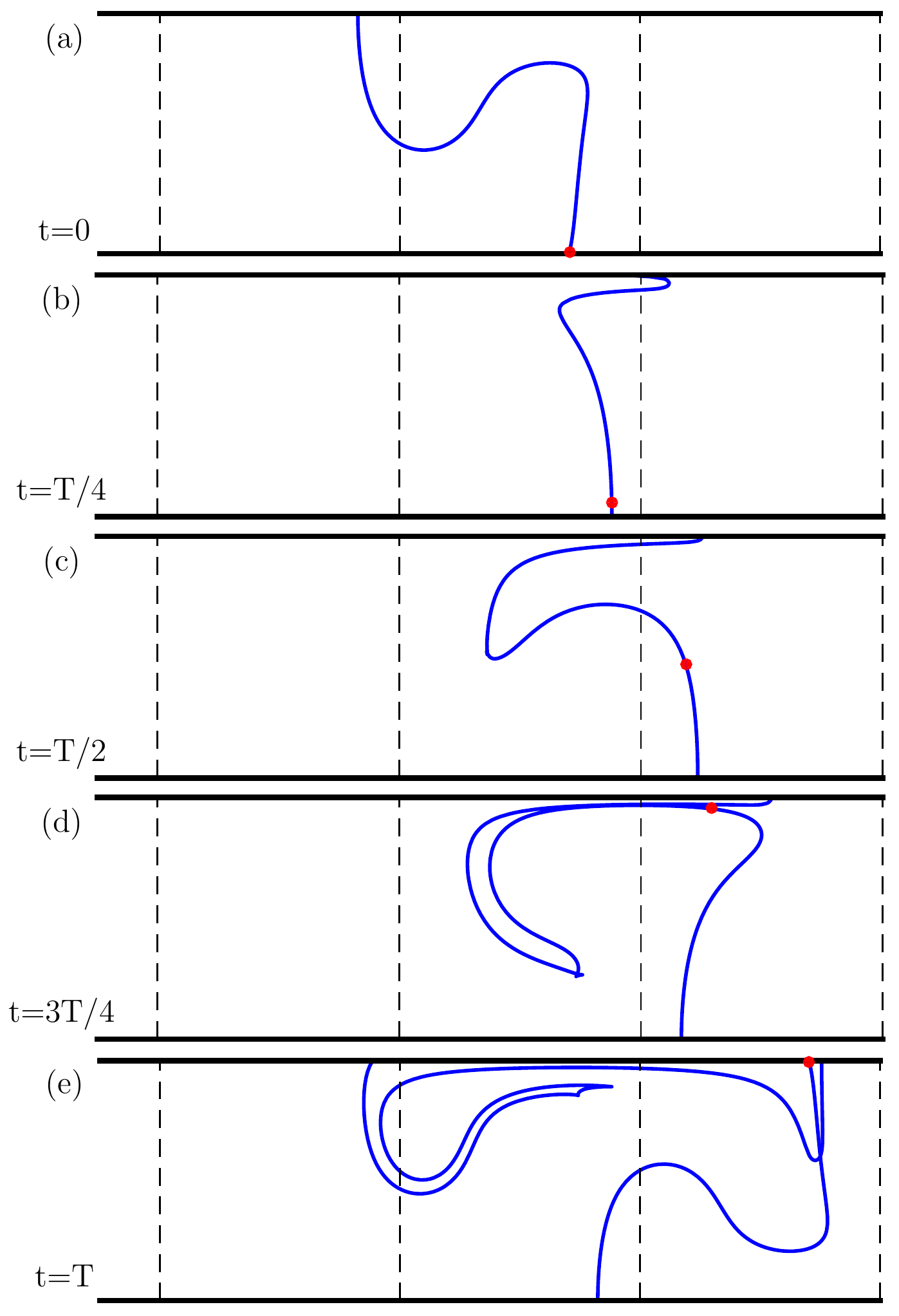}
\caption{ The mechanism by which the (1,2) mode-locked front loses its domanince with $v_0 = 0.0873$. (a) At $t = 0$ the RPO sits near the channel boundary. (b,c) The RPO moves upward while the front develops a long ``finger'' along the top boundary. (d) The RPO moves up, colliding with the ``finger''. Oppositely oriented fronts can interpenetrate. (e) Final frame shows a bounding front with multiple components. This RPO is no longer on the bounding front, and therefore not the RPO associated with the dominant mode-locking type.}
\label{fig:Global Bifurcation}
\end{figure}

\acknowledgements
The present work was supported by the US National Science Foundation under grants PHY-0748828 and CMMI-1201236.

\appendix
\section{\label{sec:level6}Numerical computation of RPOs}

We employ the following method for finding RPOs. For a given value of $v_0$, a reaction is stimulated and evolved forward until there appears to be a consistent pattern in the reaction front of some mode-locking type. We have shown in Fig.~\ref{fig:Front_Convergence} that the convergence to the mode-locked front is fast, thus after running a simulation we can infer the mode-locking type by plotting the results similar to those shown in Fig.~\ref{fig:ML_all}. To confirm our intuition we can take a front(s) and the suspected mode-locking type $(N,M)$, and apply  (Eq.~\ref{r4}) to ensure the front(s) is invariant. Finally, we employ a ``multishooting" approach, with $M$ initial guesses, and Newton's method to find the RPOs for the given value of parameters. The multishooting approach and the way in which we choose the initial points are outlined below. 

\subsection{\label{sec:level1}Multishooting approach} 
For an assumed mode-locking type $(N,M)$, we define a map $\mathbf{G}_{N,M}$ which operates on a set of $M$ points.
This composite map first applies $F$ to each of the $M$ points.
It then shifts only the $M$th point backward by $2N$ vortices.
Finally, the indices of the $M$ points is cyclically permuted ($1 \to 2, \ldots, M \to 1$).
While the RPO is a periodic orbit of $F_{N,M}$, it is a fixed point of $\mathbf{G}_{N,M}$.
We demonstrate this graphically for type (2,3) mode locking in Fig.~\ref{fig:23ML_seeds}. 
Each initial guess (diamond) is mapped forward for one period (circle).
The last point is then shifted back by $2N=6$ vortices and the ordering of the points is cyclically permuted.
The RPO is now a fixed point of $\mathbf{G}_{2,3}$.

Finally, to find the RPOs, we consider the difference between inputs $\mathbf{x}$ and outputs $\mathbf{G}_{N,M}(\mathbf{x})$.
Using Newton's Method, a standard fixed point finding algorithm, we determine the RPO.

\begin{figure*}
\includegraphics[width=\linewidth]{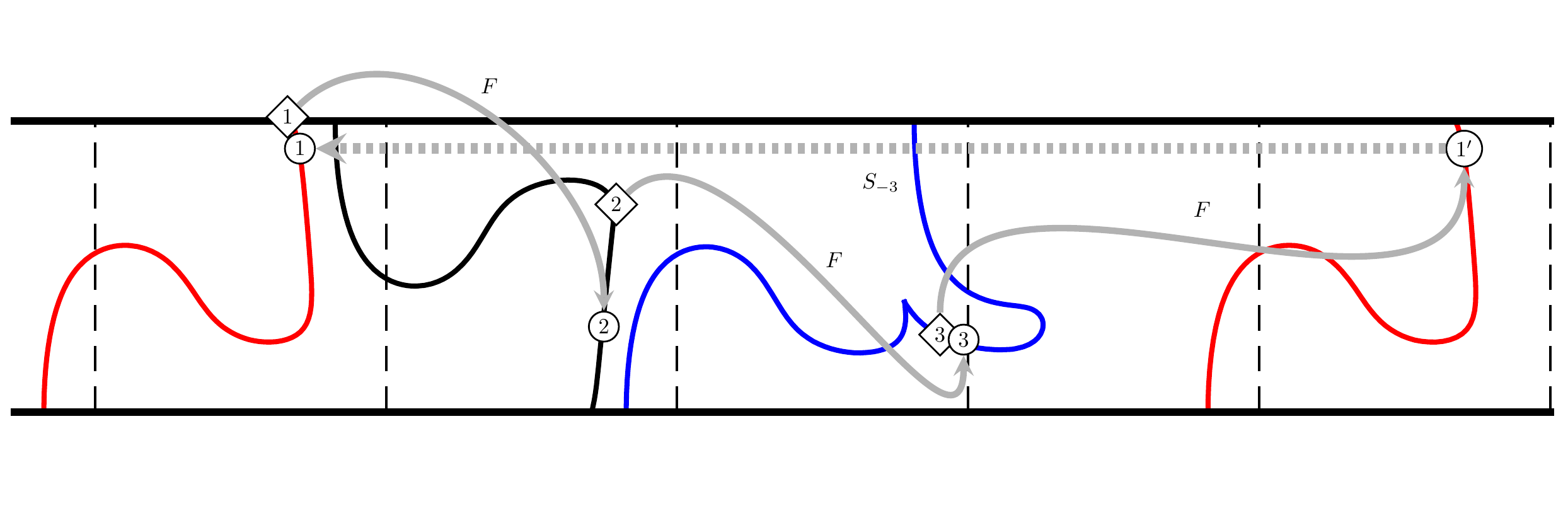}
\caption{Multishooting: The diamonds represent the seeds. These points are mapped forward under $F$ with the assumed mode-locking type, $(2,3)$ in this example, resulting in the circles. The last point is then shifted back by $2N$ vortices. Finally to find the RPOs we take the difference between the initial and mapped forward points and use Newton's method to find the roots of this difference map.}
\label{fig:23ML_seeds}
\end{figure*}

 \subsection{\label{sec:level1}Choosing an initial seed} 
The chaotic behavior of the system dictates that we take care in choosing our seeds.
For example, it was found that for $(3,5)$ mode-locking, the above method only converged when the seed was already within $1\e{-4}$ in each of the nine dimensions.
We employ the following algorithm:
First, we evolve an arbitrary front until it has converged to a mode-locked front (blue spanning front in Fig.~\ref{fig:Seed Finding}).
This front is then parameterized using its (2D) Euclidean distance and resampled at a high density (using about 10,000 points). 
Maintaining this same parameterization, it is evolved forward for several more periods. (Here, two is sufficient.)
Due to the extreme stretching, all points on the last front (black front in Fig.~\ref{fig:Seed Finding}) have the same parameter value to within $1\e{-5}$ of each other. 
We then used any one of the parameter values on the last front (black front in Fig.~\ref{fig:Seed Finding}) to determine where the point lies on the first parameterized front (blue front in Fig.~\ref{fig:Seed Finding}).
For the $(2,3)$ mode-locking type, we simply used the point on the first front (blue) whose parameter value was closest to the chosen point on the last front (black). 
For the higher order $(3,5)$ type, it was necessary to interpolate a point on the first parameterized front with the chosen parameter value from the last front.

This process is repeated for each successive point of the RPO for the assumed mode locking type.With these seeds we utilized the multishooting method described above to find the RPOs. This method of choosing seeds provided the highest chance of convergence of Newton's method.

\begin{figure}
\includegraphics[width=\linewidth]{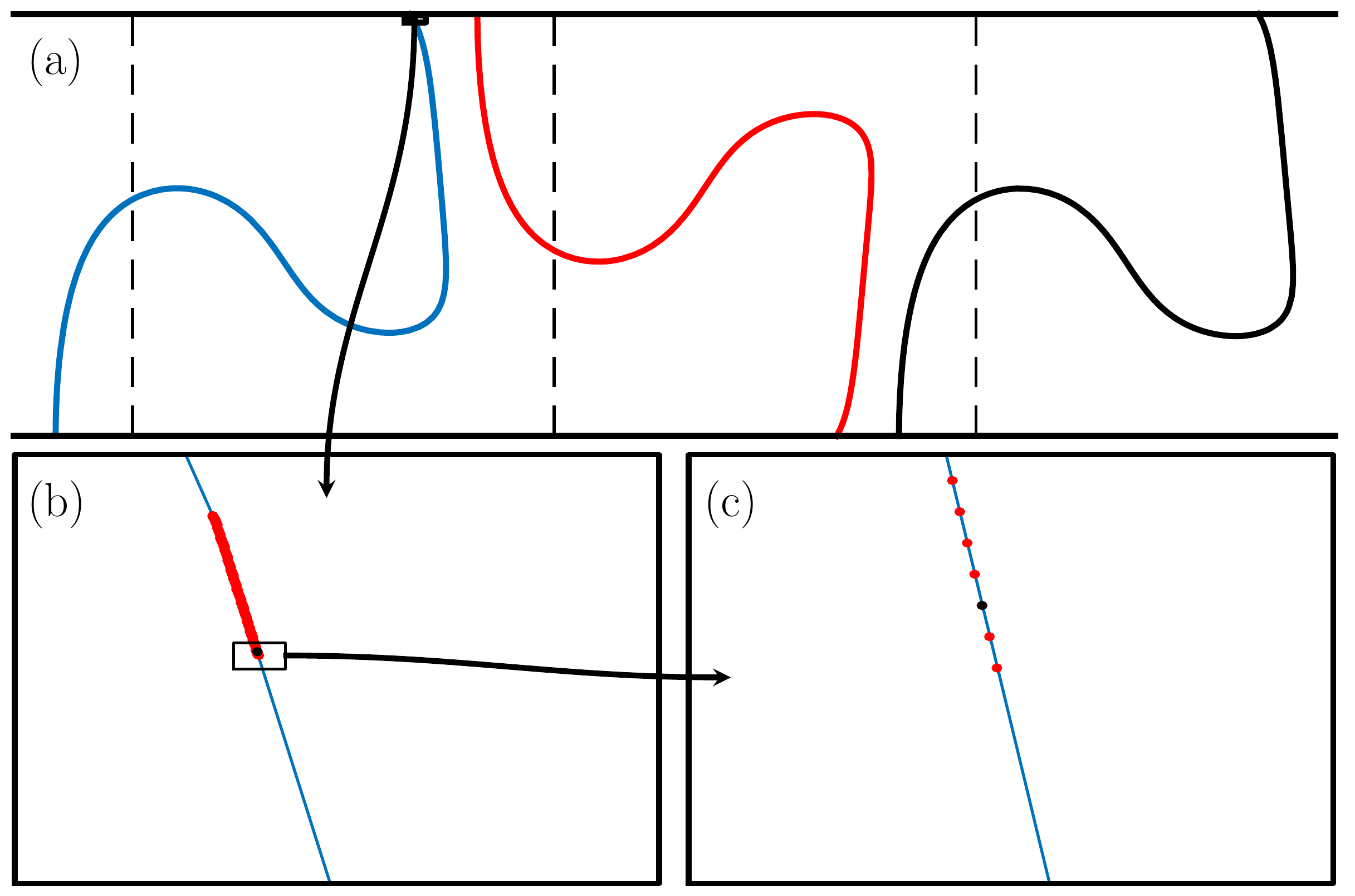}
\caption{Seed Finding: (a) Shows mode-locking of type (1,2). The blue front is mapped forward for two periods resulting in the red and black fronts, respectively. The rectangle on the blue front in (a) encompasses the points that, once mapped forward, comprise the red and black fronts.  (b) shows the zoomed in region in (a). Here a new rectangle encompasses the region where the black front originates. (c) shows the zoomed in region in (b). For finding good initial guesses for Newtons method, the black point remaining in (c) is a seed for the RPO.}
\label{fig:Seed Finding}
\end{figure}

\bibliography{mybib}

\end{document}